\documentstyle[preprint,aps,amsfonts]{revtex}

\begin{document}

\draft

\title{Collective and single-particle excitations of a 
trapped Bose gas}

\author{F. Dalfovo,$^1$ S. Giorgini,$^2$ M. Guilleumas,$^1$
L. Pitaevskii,$^{3,4}$ and S. Stringari$^1$}

\address{$^1$ Dipartimento di Fisica, Universit\`a di Trento, and \\
Istituto Nazionale Fisica della Materia,
I-38050 Povo, Italy}

\address{$^2$ European Centre for Theoretical Studies in Nuclear
Physics and Related Areas \\
Villa Tambosi, Strada delle Tabarelle 286, I-38050
Villazzano, Italy} 

\address{$^3$ Department of Physics, TECHNION, Haifa 32000, Israel}

\address{$^4$ Kapitza Institute for Physical Problems, ul. Kosygina 2,
117334 Moscow}

\date{June 30, 1997}

\maketitle

\begin{abstract}
The density of states of a Bose-condensed gas confined in a harmonic
trap is investigated. The predictions of Bogoliubov theory are compared
with the ones of Hartree-Fock theory and of the hydrodynamic model. We
show that the Hartree-Fock scheme provides an excellent description
of the excitation spectrum in a wide range of energy, revealing a 
major role played by single-particle excitations in these confined 
systems. The crossover from the hydrodynamic regime, holding at low
energies, to the independent particle regime is explicitly explored 
by studying the frequency of the surface mode as a function of their 
angular momentum. The applicability of the semiclassical 
approximation for the excited states is also discussed. We show that 
the semiclassical approach provides simple and accurate formulae 
for the density of states and the quantum depletion of the
condensate. 
\end{abstract}


\narrowtext

\section{INTRODUCTION}
\label{sec:intro}

The collective modes of a Bose condensed gas confined by an external
potential  have been the object of extensive work in the last months.
The successful  agreement between experimental results 
\cite{Jila,MIT} and theoretical predictions
\cite{Stringari,Edwards,Singh,Esry,Zaremba,You}
for the collective frequencies  at low temperature
has  stimulated an intensive research activity. Though only the modes
with low  multipolarity and frequency have been detected in
experiments, the  excitations at higher energy and angular 
momentum are also very important
because they determine the  statistical behavior of the system,
including thermodynamics, transport phenomena and
superfluid effects. 

The excited states at high energy are expected to have single-particle 
nature.  However, the transition from collective (phonon-like) to 
single-particle excitations in a inhomogeneous system can differ 
significantly from the case of a uniform Bose gas. In fact, 
the  presence of a surface allows for the occurrence of single-particle 
states even at low energy (lower than the chemical potential). These
states, of low energy but high multipolarity,  are localized near 
the surface, where the condensate density
becomes small. This behavior represents a peculiar and interesting 
feature of these confined systems; in a uniform Bose gas, in fact, only 
phonons are present at low energy.  In a recent paper \cite{Giorgini} 
we have already pointed out the effects of these single-particle states 
on the  thermodynamic properties of the trapped  gases. 

In the present work we solve the equations for the excited states
of a weakly interacting  gas in a spherical trap at zero 
temperature within Bogoliubov  theory.  
The main purpose is to investigate  the collective 
(phonon-like) and single-particle character of the elementary 
excitations.  This is accomplished by calculating key quantities, 
such as the  density of states, the frequency of the surface 
modes and the quantum depletion of the condensate, and by 
comparing the predictions of Bogoliubov theory with the ones of 
different approximations, like  Hartree-Fock theory 
and the hydrodynamic model.  Finally we check the accuracy of 
the semiclassical approximation and we show that it provides 
simple and useful formulae for both the density of states and the 
quantum depletion of the condensate. 

\section{Bogoliubov theory} 
\label{sec:bogoliubov}

The elementary excitations of a degenerate Bose gas are  associated
with the fluctuations of the condensate. At low temperature they are
described by the time  dependent Gross-Pitaevskii (GP) equation for the
order parameter \cite{GP}:
\begin{equation}
 i\hbar {\partial \over \partial t} \Psi ({\bf r},t) =
\left( - { \hbar^2 \nabla^2 \over 2m} + V_{ext}({\bf r}) 
+ g \mid \!\Psi({\bf r},t) \!\mid^2 \right) \Psi({\bf r},t)  \; , 
\label{TDGP}
\end{equation}
where $\int \! d{\bf r} |\Psi|^2= N$ is the number of atoms in 
the condensate. At zero 
temperature $N$ coincides with the total number of atoms, except 
for a very small difference $\delta N \ll N$ due to the 
quantum depletion of the condensate.  The coupling constant 
$g$ is proportional to the  $s$-wave scattering  length $a$ through 
$g=4\pi \hbar^2 a/m$. In the present work we will  discuss the case 
of positive scattering  length, $a>0$, as in the  experiments with 
Rubidium and Sodium, but the same formalism can be also applied to 
systems with negative scattering length.  
The trap is included through $V_{ext}$, which is chosen here in the 
form of an isotropic harmonic potential:  $V_{ext}(r)= (1/2) m 
\omega_{_{HO}}^2 r^2$. The harmonic trap provides also a typical 
length scale for the system,  $a_{HO}= (\hbar/m\omega_{HO})^{1/2}$.  
Actually, the experimental  traps have cylindrical 
symmetry, with different radial and axial frequencies, but the choice 
of a spherical trap, as we will discuss later, is not expected to
affect  the main conclusions of the present work, while reducing greatly
the numerical effort. 

The normal modes of the condensate can be found by linearizing the GP 
equations, i.e., looking for solutions of the form
\begin{equation}
\Psi({\bf r},t) = e^{-i\mu t/\hbar} \left[ \Psi_0 
({\bf r}) + u({\bf r}) e^{-i \omega t} + v^*({\bf r}) e^{i \omega t}
\right] 
\label{linearized}
\end{equation}
where $\mu$ is the chemical potential and functions $u$ and $v$ are 
the ``particle" and ``hole" components characterizing the Bogoliubov 
transformations. After inserting in (\ref{TDGP}) and retaining 
terms up to first order in $u$ and $v$, one finds three equations. 
The first one is the nonlinear equation for the 
order parameter of the ground state \cite{GP}, 
\begin{equation}
\left[ H_0 + g \Psi_0^2 ({\bf r})  \right] \Psi_0({\bf r})
= \mu   \Psi_0({\bf r}) \, , 
\label{groundstate}
\end{equation}
where $H_0= - (\hbar^2/2m) \nabla^2 +  V_{ext}({\bf r})$;  while 
$u({\bf r})$  and $v({\bf r})$ obey the following coupled equation:  
\begin{eqnarray}
  \hbar \omega u({\bf r}) &=& [ H_0 - \mu + 2 g \Psi_0^2]  u ({\bf r})
+ g  \Psi_0^2 v ({\bf r})  
\label{coupled1}
\\
- \hbar \omega v({\bf r}) &=& [ H_0 - \mu + 2 g \Psi_0^2]  v ({\bf r})
+ g  \Psi_0^2 u ({\bf r}) \; . 
\label{coupled2}
\end{eqnarray}
Numerical solutions of these equations have been recently found by 
different authors \cite{Edwards,Singh,Esry,Zaremba,You}. In the 
present work, we use them to calculate the density of states, the
frequency of the surface modes, the quantum depletion of the 
condensate, in order to clarify the different role played by
excitations having collective and single-particle character.  
 
When the dimensionless parameter $N a/a_{_{HO}}$ is
large, the kinetic energy term in the ground state equation 
(\ref{groundstate}) becomes negligible with respect to the 
mean field term and one gets the Thomas-Fermi approximation:
\begin{equation}
\Psi^{TF}_0 ({\bf r}) 
= \left( {\mu^{TF} - V_{ext} ({\bf r}) \over g  }
\right)^{1/2}
\label{TF}
\end{equation}
with 
\begin{equation}
\mu^{TF} = { \hbar \omega_{HO} \over 2}  \left( 15 {N a 
\over a_{_{HO}}} \right)^{2/5} \; .  
\label{muTF}
\end{equation}
In the same limit  the equations of motion (\ref{coupled1}-\ref{coupled2})  
coincide with the equations of the hydrodynamics (HD) of  superfluids 
\cite{Stringari,Fetter,Wu}. 
In the spherical case their eigenfrequencies 
take the analytic form \cite{Stringari}
\begin{equation}
\omega (n,\ell) = \omega_{HO} (2 n^2 + 2 n \ell + 3 n + \ell)^{1/2} 
\; ,
\label{HDspectrum}
\end{equation}
where $\ell$ and $n$ are the angular momentum quantum number and the 
number of nodes in the radial solution, respectively. The 
deviations from the  predictions of the noninteracting  harmonic 
oscillator (HO) model, 
\begin{equation}
\omega (n,\ell) = \omega_{HO} (2 n + \ell) \; , 
\label{HOspectrum}
\end{equation}
point out the effects of two-body interactions.  These are 
particularly important for the so called ``surface" modes ($n=0$), where  
the HO prediction $\omega = \ell \omega_{H0}$ is significantly 
lowered to the hydrodynamic  value $\omega =\sqrt{\ell} 
\ \omega_{H0}$. In general the HD prediction turns out to be 
very accurate for the low energy excitations of the system, while 
the ideal gas prediction is expected to be valid in the
opposite case of high excitation energies. The exact solutions of the 
equations (\ref{coupled1}-\ref{coupled2}) provide the correct 
interpolation between the two limiting regimes. 

A typical spectrum obtained from equations 
(\ref{coupled1}-\ref{coupled2}) is given in the upper part of
 Fig.~\ref{fig:spectrum}
for a  gas of $N=10000$ atoms of Rubidium  (scattering 
length $a=110 a_0$, where $a_0$ is the Bohr radius). For the spherical 
trap we have chosen the frequency $\omega_{HO}=2\pi \nu_{HO}= 2\pi 187$Hz,
which is the average $\omega_{HO}= (\omega_x \omega_y \omega_z)^{1/3}$
of the axial and radial frequencies of Ref.~\cite{Jila}. 
It corresponds to 
the oscillator length $a_{HO}=0.791 \times 10^{-4}$ cm. Energy is 
given in units $\hbar \omega_{HO}$ and the chemical potential is 
$8.41$ in  these units. The vertical bars have length 
$(2 \ell +1)$, so that the angular momentum of each state can be
inferred from the figure. One clearly sees that, at energy much larger
than the chemical potential, the excited states tend to be grouped into 
levels $\hbar \omega_{HO}$ apart, as in the noninteracting HO model. 
Conversely, the energy of the lowest modes is close to the prediction
of the HD equations \cite{Stringari}. For instance, the 
lowest $\ell=2$ and $\ell=0$ modes differ by less than $3$\% from 
the hydrodynamic values $\sqrt{2}$ and $\sqrt{5}$, respectively. 

\section{Density of states}
\label{sec:densityofstates}

Once the spectrum of excited states is calculated, one can count the 
number of states below a given energy $\epsilon$: 
\begin{equation} 
{ \frak{N} }  (\epsilon) = \sum_{\hbar \omega < \epsilon } (2 \ell +1 ) \; . 
\label{numberofstates}
\end{equation}
The density of states, $g(\epsilon)$ is simply the derivative 
$d{ \frak{N} } (\epsilon)/d\epsilon$. Note that the quantity 
${ \frak{N} } (\epsilon)$ is well defined even for a discretized 
spectrum, while the density of states implies averaging the number of
states within small but finite energy intervals. 

In Fig.~\ref{fig:N(E)} we show  the quantity ${ \frak{N} }  (\epsilon)$ 
obtained by counting the levels in the spectrum of 
Fig.~\ref{fig:spectrum}.
For comparison the results of the noninteracting harmonic 
oscillator (\ref{HOspectrum}) and of the hydrodynamic model 
(\ref{HDspectrum}) are also shown. The effects 
of the repulsive interatomic forces are clearly responsible for 
an  enhancement of  the density of states with respect to the ideal 
gas. However ${ \frak{N} } (\epsilon)$ remains well below the HD 
approximation, 
the latter being soon inadequate as $\epsilon$ increases. Indeed  
hydrodynamic theory accounts for collective phenomena and provides 
an excellent description of the low-lying elementary excitations of 
the  system \cite{Stringari,You},  but completely ignores single-particle 
effects. This is exactly the opposite of what  Hartree-Fock (HF) theory
does. For this reason it is interesting to compare the results of
Bogoliubov theory  with
the predictions of HF theory in which one determines the 
eigenstates of the single-particle Hamiltonian \cite{HF,noteHF}
\begin{equation}
H_{HF} = - (\hbar^2/2m) \nabla^2 +  V_{ext}({\bf r}) -\mu + 
2 g  \Psi_0^2 ({\bf r}) \; .
\label{HFhamiltonian}
\end{equation}
The lowest eigenstates of the
HF Hamiltonian are expected to be localized near the surface of the 
condensate. To understand better this point, let us take the large
$N$ limit. In this case one can use the Thomas-Fermi approximation 
(\ref{TF}) for the ground state density. The HF Hamiltonian then 
takes the simple form
\begin{equation}
H_{HF} =  - (\hbar^2/2m) \nabla^2  + {1\over 2} m \omega^2_{HO}
|r^2 -R^2| \; ,
\label{largeNHF}
\end{equation}
where $R=[2 \mu^{TF}/(m\omega^2_{HO})]^{1/2}$ is the classical radius
of the condensate. The HF potential has a pronounced minimum at $R$.
This potential well near the boundary persists in the HF Hamiltonian
even for smaller values of $N$.  

It is worth stressing that, in general, Hartree-Fock theory is
expected to be correct for energies larger than the chemical potential.
For these trapped bosons, however,  it accounts also for the low energy   
excitations close to the boundary, where the density of the condensate 
is small. This can be seen in Fig.~\ref{fig:spectrum}, where the 
Hartree-Fock and Bogoliubov spectra are compared. One notes that
the two spectra are rather similar even  below the chemical 
potential ($\mu=8.41$ in this case) except for the excitations 
having lowest angular momenta, i.e., shortest bars in the figure. 
Those collective excitations are  instead correctly reproduced by the 
hydrodynamic model. The corresponding Hartree-Fock prediction for the 
quantity ${ \frak{N} } (\epsilon)$ is also given in Fig.~\ref{fig:N(E)} as
a solid line. The agreement with the results of 
equations (\ref{coupled1}-\ref{coupled2}) is remarkable in the 
whole range of energy.  The figure shows the case of $10000$ atoms but
a similar agreement is found  for all values 
of $N$ relevant for the experiments ($N \simeq 10^3 \div 10^7$).  
The above  behavior represents a major difference with  
respect to the case of a homogeneous Bose gas where phonons 
have a crucial effect on the density of states. 

\section{Surface excitations}
\label{sec:surface} 

In order to better understand the transition from the collective
to the  single particle regime, we have explored in details the 
evolution of the excitation energy for the surface modes ($n=0$) as a 
function of their angular momentum  $\ell$, as predicted by the
solution of the  equations (\ref{coupled1}-\ref{coupled2}). 
As already 
mentioned, the effects of the interaction are particularly 
important  for such modes and are responsible for a significant lowering 
of  their frequency. In Fig.~\ref{fig:omegaell} we plot the 
quantity $\omega_{\ell}/\ell$, in units $\omega_{HO}$, for different 
numbers of atoms in the same trap. This ratio has an important physical 
meaning because, according to Landau's criterium for superfluidity, 
it provides the rotational frequency at which the $\ell$-th surface 
excitation becomes unstable. The hydrodynamic prediction 
(\ref{HDspectrum}) is shown as a dashed line, the curve going 
asymptotically to zero for large $\ell$.  The figure shows  that 
the Bogoliubov states first follow the HD curve, but,  rather soon, 
they deviate from it,  approaching asymptotically the non 
interacting value $\omega_{\ell}/\ell= \omega_{HO}$. The 
deviation from HD takes place at larger values of $\ell$ if $N$ is
increased, revealing that the HD approximation becomes applicable to a
larger number of states in this limit.  

A  simple estimate of the
typical value of $\ell$ at which the HD picture starts failing, one
can take \cite{Pethick} $\ell_{c} \simeq R p_{c}$ where $R$ is the 
radius of the condensate, proportional to $N^{1/5}$, and 
$p_{c}$ is of the  order of the inverse of the surface thickness 
$d = [a_{HO}^4/(2R)]^{1/3}$ \cite{Baym,boundary,PethickPRA}. 
For larger values  of $\ell$, the wavelength of the excitations
becomes shorter than $d$ and one explores microscopic details   
of the boundary which can not be described by the Thomas-Fermi 
approximation (\ref{TF}) and by the HD equations. This yields 
$l_{c} \propto (R/a_{HO})^{4/3} \propto N^{4/15}$, corresponding 
to an excitation energy  $\hbar \omega \propto N^{2/15}$, smaller
than the chemical potential, which instead behaves as
$N^{2/5}$. This  explains why  the crossover from the HD to 
the single particle regime takes place at energies 
smaller than $\mu$.  

For each value of $N$, the curves in Fig.~\ref{fig:omegaell} 
exhibit minima and one can define a critical frequency as 
$\Omega_c =\hbox{min} (\omega_{\ell}/\ell)$. For rotational 
frequencies larger than 
$\Omega_c$ the surface excitations become unstable.  It is 
interesting to compare this value with the  critical frequency 
needed to generate a vortex \cite{Dalfovo}. 
This is done in Fig.~\ref{fig:omegac} 
where we compare the two critical frequencies as a function of $N$. 
We find that the lowest instability is always associated with the
creation of a vortex. Note however that in order to generate a vortex
one needs to transfer to the system a  huge angular momentum (equal to
$N\hbar$), which is much higher than the value $\ell$ required to
create a surface excitation.

\section{Semiclassical approximation and scaling behavior} 
\label{sec:scaling}

A good approximation for the density of states can be obtained 
by solving equations (\ref{coupled1}-\ref{coupled2}) in 
the semiclassical approximation \cite{Giorgini,semiclassical,Huang}.  
In this approximation, which 
is expected to hold for excitation energies much larger than the 
oscillator energy $\hbar\omega_{HO}$, the quantity ${ \frak{N} } (\epsilon)$ 
is a continuous function of $\epsilon$ defined by
\begin{equation} 
{ \frak{N} }  (\epsilon) = \int_0^\epsilon \! d\epsilon' \int \!
{ d{\bf r}d{\bf p} \over (2\pi \hbar)^3} \ \delta(\epsilon' -\epsilon
({\bf r},{\bf p})) 
\label{semiclassical} 
\end{equation}
where 
\begin{equation} 
\epsilon({\bf r},{\bf p}) = \left[ \left( {p^2 \over 2m} + V_{ext}
({\bf r}) - \mu + 2 g n_0 ({\bf r}) \right)^2 - g^2 n^2_0 ({\bf r})
\right]^{1/2}  
\label{epsilonsemiclassical}
\end{equation}
corresponds to the semiclassical dispersion law. Here the quantity 
$n_0({\bf r}) = \Psi^2_0 ({\bf r})$ is the condensate density. 
In Fig.~\ref{fig:semiclassical}a we compare the semiclassical 
result for ${ \frak{N} }  (\epsilon)/N$ (solid line) with the one 
obtained from equations (\ref{coupled1}-\ref{coupled2}) 
(squares)  for $10^4$  atoms of Rubidium in the same trap of 
Fig.~\ref{fig:N(E)}. 
Here the energy is given in units $k_B T_c = \hbar \omega_{HO} 
[N/\zeta(3)]^{1/3}$,  which is the critical
temperature for an ideal Bose gas in a harmonic trap; the value
for $10^4$ atoms is $k_B T_c = 20.26 \hbar\omega_{HO}$ as shown 
also in Fig.~\ref{fig:spectrum}.
The accuracy  of  the semiclassical approximation turns out to 
be very high also for relatively low values of $\epsilon$. 

The use of the  semiclassical approximation allows one to 
carry out the analysis of the
density of states in a  systematic way and to exploit the dependence 
on the  relevant parameters of the system. In fact, when the number 
of atoms in the condensate is large  enough to make  the Thomas-Fermi 
approximation (\ref{TF})  accurate, the statistical 
properties of  the system can be expressed in terms of a 
single scaling parameter $\eta$ given by the ratio \cite{scaling}  
\begin{equation}
\eta = {\mu^{TF} \over k_B T_c} = 1.57 \left( { N^{1/6} a \over
a_{HO} } \right)^{2/5}  
\label{eta}
\end{equation}
between the chemical potential (\ref{muTF}), calculated at zero 
temperature in the  Thomas-Fermi approximation, and the critical 
temperature $k_B T_c$.   The ratio $\eta$ 
depends on the deformation of the trap only through 
the geometrical average of the oscillator  frequencies 
$\omega_{HO} = (\omega_x \omega_y \omega_z)^{1/3}$.  The
parameters used in the calculation of Fig.~\ref{fig:semiclassical}a 
correspond  to a spherical trap with $\eta = 0.407$.  As pointed 
out in Ref.~\cite{scaling}, quite different experimental 
conditions (shape  of the trap,  value of $N$,  etc.) can 
correspond to very similar values  of $\eta$. In terms of the  
scaling parameter $\eta$ and the 
dimensionless energy  $\tilde{\epsilon}= \epsilon/(k_b T_c)$,
the number of states ${ \frak{N} } (\tilde{\epsilon})$ predicted by the 
Bogoliubov semiclassical theory becomes
\begin{eqnarray}
{ { \frak{N} } (\tilde{\epsilon}) \over N} =  
\int_0^{\tilde{\epsilon}} \! &d& \tilde{\epsilon}'  {4 \over \pi 
\zeta(3) } \int_0^1 \! dx \ \sqrt{1-x} \nonumber \\ 
& \times & \left[ \tilde{\epsilon}' \eta 
{ \sqrt{ [x^2 +(\tilde{\epsilon}')^2/\eta^2]^{1/2} - x } \over 
\sqrt{ x^2 + (\tilde{\epsilon}')^2/\eta^2} }  + 
(\tilde{\epsilon}')^2 \sqrt{ x + \eta/\tilde{\epsilon}' } \right] \; . 
\label{ratio}
\end{eqnarray}
This result has been obtained by using the Thomas-Fermi 
approximation (\ref{TF}) for the condensate density in 
(\ref{semiclassical}-\ref{epsilonsemiclassical});  
this allows one to split the space integral  into an
{\it inside region} (first term in the square bracket) and an
{\it outside region} (second term).  
It is worth stressing that equation (\ref{ratio}), which are 
expected to hold in the scaling regime $Na/a_{HO}\gg 1$, provide a 
very good estimate of the semiclassical expression 
(\ref{semiclassical}) 
even for relatively small $N$. For instance, the two predictions are 
indistinguishable in Fig~\ref{fig:semiclassical}a, being represented by 
the same solid line. In Fig.~\ref{fig:semiclassical}b we show the predictions 
for the density of states given by the  semiclassical approximation 
(\ref{ratio}) for three different values of $\eta$. The parameters 
of the recent experiments at Jila \cite{Jila} and MIT \cite{MIT}, 
using very different traps,  correspond to $\eta$ ranging from 
$0.39$ to $0.45$. 

Expression (\ref{ratio}) can be also expanded at low  energy,  
$\epsilon \ll k_B T_c$, still compatible with the assumption $\epsilon 
\gg \hbar\omega_{HO}$. One finds the law  ${ \frak{N} } (\epsilon)/N \propto 
\epsilon^{5/2}$. This differs
from the usual $\epsilon^3$ law typical of the phonon regime, revealing 
the different behavior exhibited by these systems with respect to the 
homogeneous Bose gas.

\section{Quantum depletion}
\label{sec:depletion}

In the last part of the paper we calculate the quantum depletion of 
the condensate which, according to Bogoliubov theory, is given by 
\begin{equation}
{ \delta N \over N } = { 1 \over N } \sum_j \int \! d{\bf r} \
|v_j ({\bf r})|^2  \; , 
\label{depletion}
\end{equation}
The ``hole" components $v_j$ can be obtained by solving  equations 
(\ref{coupled1}-\ref{coupled2}). 
In semiclassical approximation \cite{Giorgini} one replaces 
the sum over all the discrete states with the 
integral over ${\bf p}$ of the function 
\begin{equation}
v^2({\bf p},{\bf r}) = { 1 \over 2 \epsilon({\bf p},{\bf r}) }
\left( {p^2 \over 2m} + V_{ext} ({\bf r}) 
- \mu +2gn_0({\bf r}) +\epsilon({\bf p},{\bf r}) \right) \; , 
\label{v2}
\end{equation}
where $\epsilon({\bf p},{\bf r})$ is the single particle energy 
(\ref{epsilonsemiclassical}). In a uniform gas this expression
yields the most famous  result $\delta N/N = (8/3) (n_0 a^3/\pi)^{1/2}$. 
In the trapped gas and in the limit $Na/a_{HO}\gg1$, where 
TF approximation holds, the 
semiclassical approximation provides the simple analytic law
\cite{Giorgini}
\begin{equation}
{ \delta N \over N } = { \eta^3 \over 6 \sqrt{2} \zeta(3) } = 0.098 
\ \eta^3 \; ,
\label{eta3}
\end{equation}
with $\eta$ given in Eq.~(\ref{eta}). 
Since the available experiment correspond to $\eta \simeq 0.4$, the 
quantum depletion turns out to be less than $1$\%, as already 
pointed out in Refs.~\cite{Singh,Zaremba}. 
In Fig.~\ref{fig:depletion} we show the quantum depletion 
for $10000$ and $50000$ atoms of Rubidium obtained by summing 
over the Bogoliubov spectrum up to a given energy 
$\epsilon$ (solid lines).  
We compare  it with the prediction of the semiclassical expression 
(\ref{v2}) (dashed lines), while the arrows indicate the asymptotic 
values (\ref{eta3}), holding in the scaling limit.   
An important result emerging from 
the numerical calculation is the very slow convergency of the 
sum (\ref{depletion}).  This is not a surprise, since also
in a homogeneous gas the convergency is slow due to the $1/p^4$ tail in
the momentum  distribution at high momenta and one has to go up
to $\epsilon = 100 \mu$ in order to saturate $90$\% of the 
sum (\ref{depletion}).  

The agreement between the quantum depletion obtained from 
the  discretized sum (\ref{depletion}) over the Bogoliubov states
and from the semiclassical approximation (\ref{v2})
is satisfying and was not obvious {\it a priori}. 
Fig.~\ref{fig:depletion} shows a discrepancy of the order of $5$\%
between the two predictions for $N=10000$, while for larger $N$ the 
two curves tend to coincide. It is worth noticing that the 
two solid lines in Fig.~\ref{fig:depletion} requires the summation
of $\int \! d{\bf r} |v_j({\bf r})|^2$ over up to $15000$ different
values of $(n,\ell)$ in the Bogoliubov spectrum; the calculation
is then much heavier than the semiclassical one. The good 
accuracy of the semiclassical approach  makes it useful in 
practical situations. This is especially true for the simple 
formula (\ref{eta3})  which includes the case of anisotropic 
traps through the averaged frequency $\omega_{HO} = (\omega_x 
\omega_y \omega_z)^{1/3}$ entering the Thomas-Fermi chemical 
potential $\mu^{TF}$ and, hence, the scaling parameter $\eta$. 

Finally, the large $N$ semiclassical formula (\ref{eta3}) shows the
rather strong dependence  of the depletion on the scattering length
parameter $a$.  If the magnetic  tuning of
the scattering length will become available, it will be possible in the
future to increase significantly the value of $\eta$ and consequently
explore Bose gases where the quantum depletion is much larger.

\section{Summary}

We have investigated the elementary excitations of a dilute Bose 
gas in harmonic trap by solving the equations of Bogoliubov theory. 
Differently from the case of a uniform gas, where phonons dominate
the system at energies of the order of, or lower 
than the chemical potential, the spectrum of the trapped gas shows 
an important particle-like behavior even at low energy. This fact 
has been here explored in detail. We have compared the results
of Bogoliubov theory for the density of states with the ones of 
Hartree-Fock theory finding a very good agreement on a wide 
range of energy. We have  studied the behavior of surface modes, 
emphasizing the crossover from the low energy regime, well described  
by the  hydrodynamic model, to  the single-particle regime. This 
crossover provides also a critical frequency associated with 
a rotational instability and we have compared this frequency with 
the one needed to create a quantized vortex.   Another important 
result emerging from our analysis is the high accuracy exhibited 
by the semiclassical  approximation for the excited states. 
Finally, we have calculated the quantum depletion of 
the condensate by summing the ``hole" component $\int\!d{\bf r} 
|v({\bf r})|^2$ over all the states in the excitation spectrum of 
Bogoliubov theory. The convergency of the sum turns out to be very 
slow, as expected by the  analogy with the case of the uniform gas. 
Again we find an excellent agreement with the predictions
of the semiclassical approximation. In the limit $Na/a_{HO}\gg 1$, the
latter provides the simple and useful formula $\delta N/N = 0.098 
\ \eta^3$, in terms of the scaling parameter $\eta= 1.57 (N^{1/6}a/
a_{HO})^{2/5}$.

\acknowledgements

We thank C.J. Pethick for useful discussions about the behavior of
the surface excitations. M.G. thanks the Ministerio de Educaci\'on 
y Ciencia (Spain) for financial support. L.P.  would like to 
acknowledge the hospitality of the Dipartimento di Fisica, Universit\`a
di Trento and the financial support of the Istituto Nazionale per la
Fisica della Materia.

\begin{figure}
\caption{Excitation spectrum of $10000$ atoms of $^{87}$Rb in a 
spherical trap with $a_{HO}=0.791 \times 10^{-4}$ cm.  The vertical 
bars have length $(2 \ell +1)$. The upper spectrum corresponds to the
numerical solution of equations 
(\protect\ref{coupled1}-\protect\ref{coupled2}); the lower one is 
the spectrum of the Hartree-Fock Hamiltonian 
(\protect\ref{HFhamiltonian}). Two energy scales (in units of  
$\hbar \omega_{HO}$) are also shown in the
figure: the chemical potential $\mu=8.41$, fixed by the solution of the 
Gross-Pitaevskii equation (\protect\ref{groundstate}), and the critical 
temperature for a noninteracting gas in the same trap, $k_B T_c =
20.26$.    }
\label{fig:spectrum}
\end{figure}

\begin{figure}
\caption{Number of states ${ \frak{N} }  (\epsilon)$ {\it vs.} energy. 
The Bogoliubov (points) and Hartree-Fock (solid line) predictions, 
obtained by counting the states in Fig.~\protect\ref{fig:spectrum},
are compared with the ones of the noninteracting harmonic 
oscillator (dashed line) and of hydrodynamic equations 
(dot-dashed line).  }
\label{fig:N(E)}
\end{figure}

\begin{figure}
\caption{Frequency (in units of $\omega_{HO}$) of the $n=0$ excited 
states as a function of their  angular momentum $\ell$ for $N$ 
atoms of $^{87}$Rb in a spherical trap with $a_{HO}=0.791 
\times 10^{-4}$ cm.  The hydrodynamic prediction is shown as 
a dashed line. }
\label{fig:omegaell}
\end{figure}

\begin{figure}
\caption{Critical rotational frequency (in units of $\omega_{HO}$)
for producing of a  quantized vortex (solid line) or surface 
states (dashed line) as a function of the number of Rubidium atoms 
in the spherical trap.  }
\label{fig:omegac}
\end{figure}

\begin{figure}
\caption{Ratio ${ \frak{N} } (\epsilon)/N$ {\it vs.} $\epsilon$, in units
$k_B T_c$. In part {\it a}, the squares correspond to counting the 
states in the Bogoliubov spectrum of Fig.~\protect\ref{fig:spectrum},
while the solid line is the corresponding semiclassical approximation 
(\protect\ref{semiclassical}). The latter is indistinguishable from 
the formula (\protect\ref{ratio}), valid in the scaling 
regime $Na/a_{HO} \gg 1$. In part {\it b}, the semiclassical 
prediction (\protect\ref{ratio}) is 
given for different values of the scaling parameter $\eta$. }
\label{fig:semiclassical}
\end{figure}

\begin{figure}
\caption{Quantum depletion for $10000$ (two lower curves) and 
$50000$ (two upper curves) atoms of $^{87}$Rb in a 
spherical trap with $a_{HO}=0.791 \times 10^{-4}$ cm. The depletion
is plotted as a function of the maximum energy considered in the 
sum (\protect\ref{depletion}). Solid lines: $|v|^2$ from 
the solution of equations 
(\protect\ref{coupled1}-\protect\ref{coupled2});
dashed lines: from the semiclassical 
approximation (\protect\ref{v2}). Arrows: asymptotic scaling 
values (\protect\ref{eta3}). } 
\label{fig:depletion}
\end{figure}


\begin{references}

\bibitem{Jila} D.\ S.\ Jin, J.\ R.\ Ensher, M.\ R.\ Matthews,
C.\ E.\ Wiemann, and E.\ A.\ Cornell,
Phys. Rev. Lett. {\bf 77}, 420 (1996).

\bibitem{MIT} M.-O.\ Mewes, M.\ R.\ Andrews,
N.\ J.\ van Druten, D.\ M.\ Kurn, D.\ S.\ Durfee, C.\ G.\ Townsend,
and W.\ Ketterle, Phys. Rev. Lett. {\bf 77}, 988 (1996)

\bibitem{Stringari} S. Stringari, Phys. Rev. Lett. {\bf 77}, 2360 (1996)

\bibitem{Edwards} M. Edwards, P.\ A.\ Ruprecht,
K.\ Burnett, R.\ J.\ Dodd, and C.\ W.\ Clark,
Phys. Rev. Lett. {\bf 77}, 1671 (1996);  P. A. Ruprecht, 
Mark Edwards, K. Burnett, and Charles W. Clark, 
Phys. Rev. A {\bf 54}, 4178 (1996) ;  M. Edwards, 
R.\ J.\ Dodd, C.\ W.\ Clark, and K. Burnett, J. Res. Natl.
Inst. Stand. Technol. {\bf 101}, 553 (1996). 

\bibitem{Singh} K.G. Singh and D.S. Rokhsar, Phys. Rev. Lett.
{\bf 77}, 1667 (1996) 

\bibitem{Esry} B.D. Esry, Phys. Rev. A {\bf 55}, 1147 (1997)

\bibitem{Zaremba} D.A.W. Hutchinson, E. Zaremba, and A. Griffin,
Phys. Rev. Lett. {\bf 78}, 1842 (1997)

\bibitem{You}  L. You, W. Hoston, and M. Lewenstein, 
Phys. Rev. A {\bf 55}, R1581 (1997)

\bibitem{Giorgini} S. Giorgini, L.P. Pitaevskii, S. Stringari,
preprint cond-mat/9704014

\bibitem{GP} L.P.~Pitaevskii, Zh. Eksp. Teor. Fiz. {\bf 40}, 646 (1961)
[Sov. Phys. JETP {\bf 13}, 451 (1961)]; E.P.~Gross, Nuovo Cimento
{\bf 20}, 454 (1961); E.P.~Gross, J. Math. Phys. {\bf 4}, 195 (1963)

\bibitem{Fetter} A.L. Fetter, Phys. Rev. A {\bf 53}, 4245 (1996);
A.L. Fetter and D. Rokhsar, preprint cond-mat/9704234

\bibitem{Wu} W.-C. Wu and A. Griffin, Phys. Rev. A {\bf 54}, 4204 (1996)

\bibitem{HF} V.V. Goldman, I.F. Silvera, and A.J. Leggett, Phys. Rev. B
{\bf 24}, 2870 (1981); D.A. Huse and E.D. Siggia, J. Low Temp. Phys. 
{\bf 46}, 137 (1982)

\bibitem{noteHF} Actually  Hamiltonian (\protect\ref{HFhamiltonian}) ignores
nonlocality effects which modify mainly the $\ell=0$ solutions of the 
Hartree-Fock theory. These states are however irrelevant for the 
determination of the density of states. 

\bibitem{Pethick} C. Pethick, private communication.

\bibitem{Baym} G.\ Baym and C.\ Pethick, Phys. Rev. Lett.
 {\bf 76}, 6 (1996).

\bibitem{boundary} F. Dalfovo, L. Pitaevskii, and S. Stringari, 
Phys. Rev. A {\bf 54}, 4213 (1996).  

\bibitem{PethickPRA} E. Lundh, C.J. Pethick, and H. Smith, 
Phys. Rev. A {\bf 55}, 2126 (1997)

\bibitem{Dalfovo} F. Dalfovo and S. Stringari, Phys. Rev. A {\bf 53},
2477 (1996)

\bibitem{semiclassical} S. Giorgini, L. Pitaevskii, and S. Stringari,
Phys. Rev. A {\bf 54}, 4633 (1996)

\bibitem{Huang} E. Timmermans, P. Tommasini, and K. Huang, Phys. Rev.
A {\bf 55}, 3645 (1997)

\bibitem{scaling} S. Giorgini, L. Pitaevskii, and S. Stringari, 
Phys. Rev. Lett. {\bf 78}, 3987 (1997) 


\end{references}
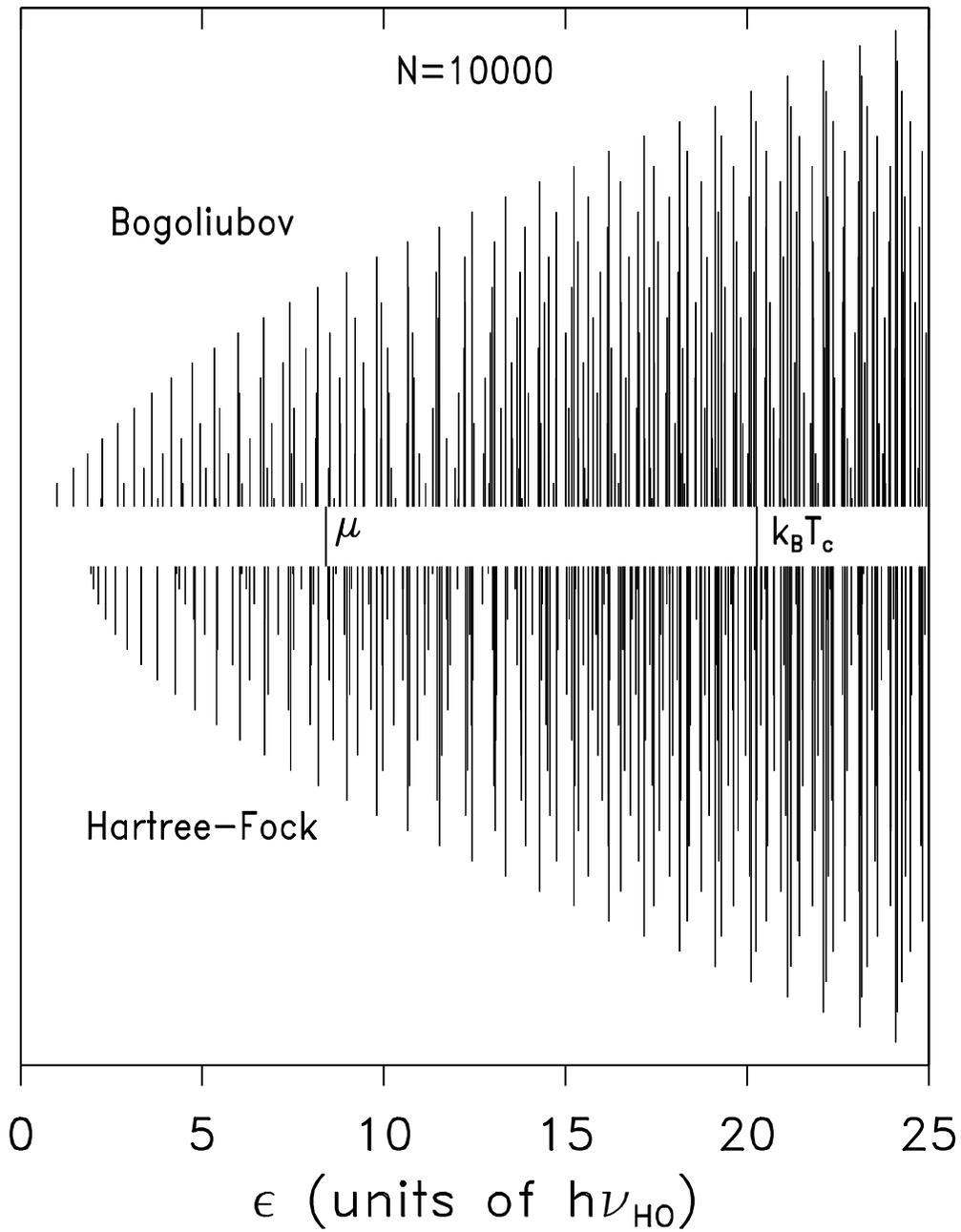
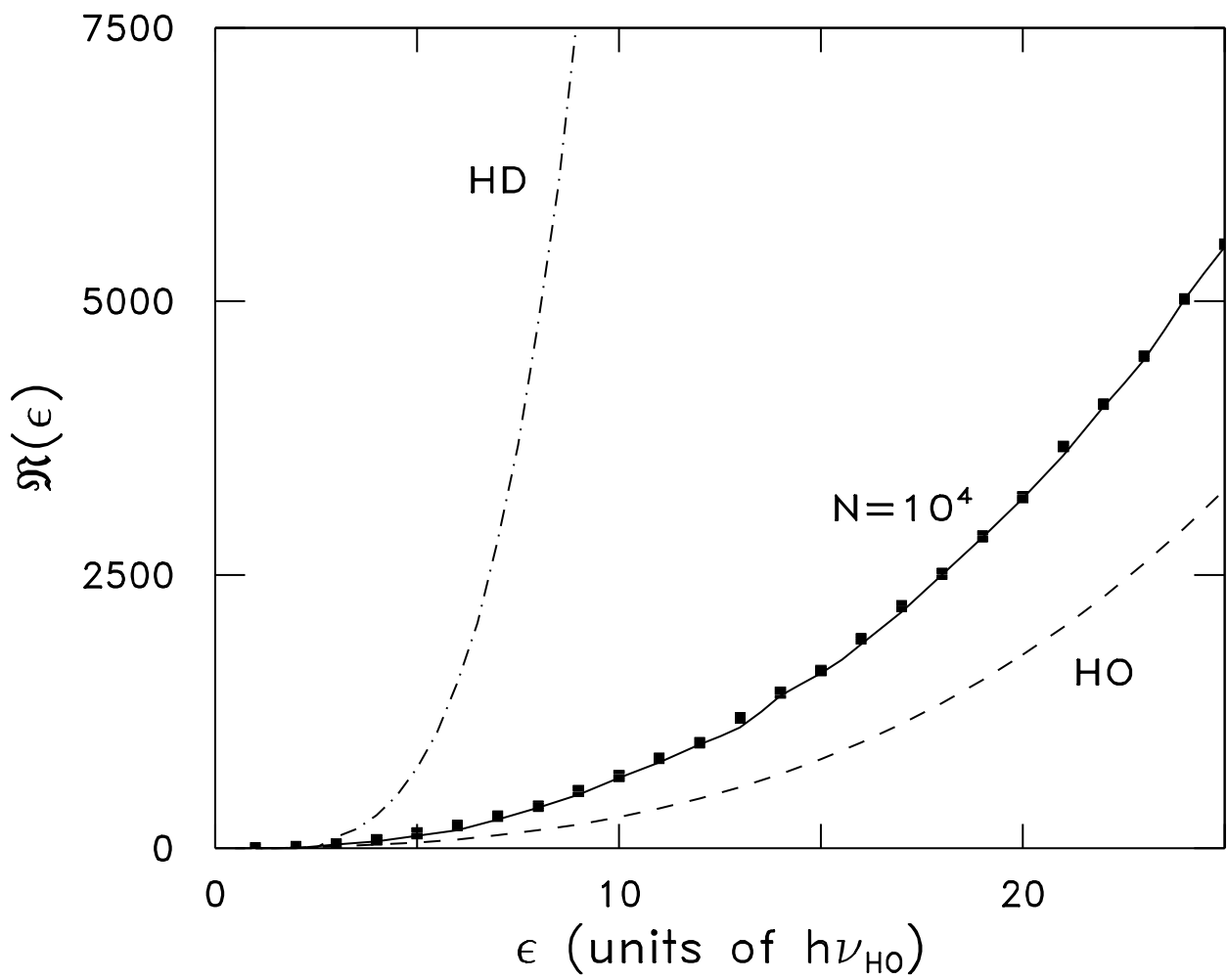
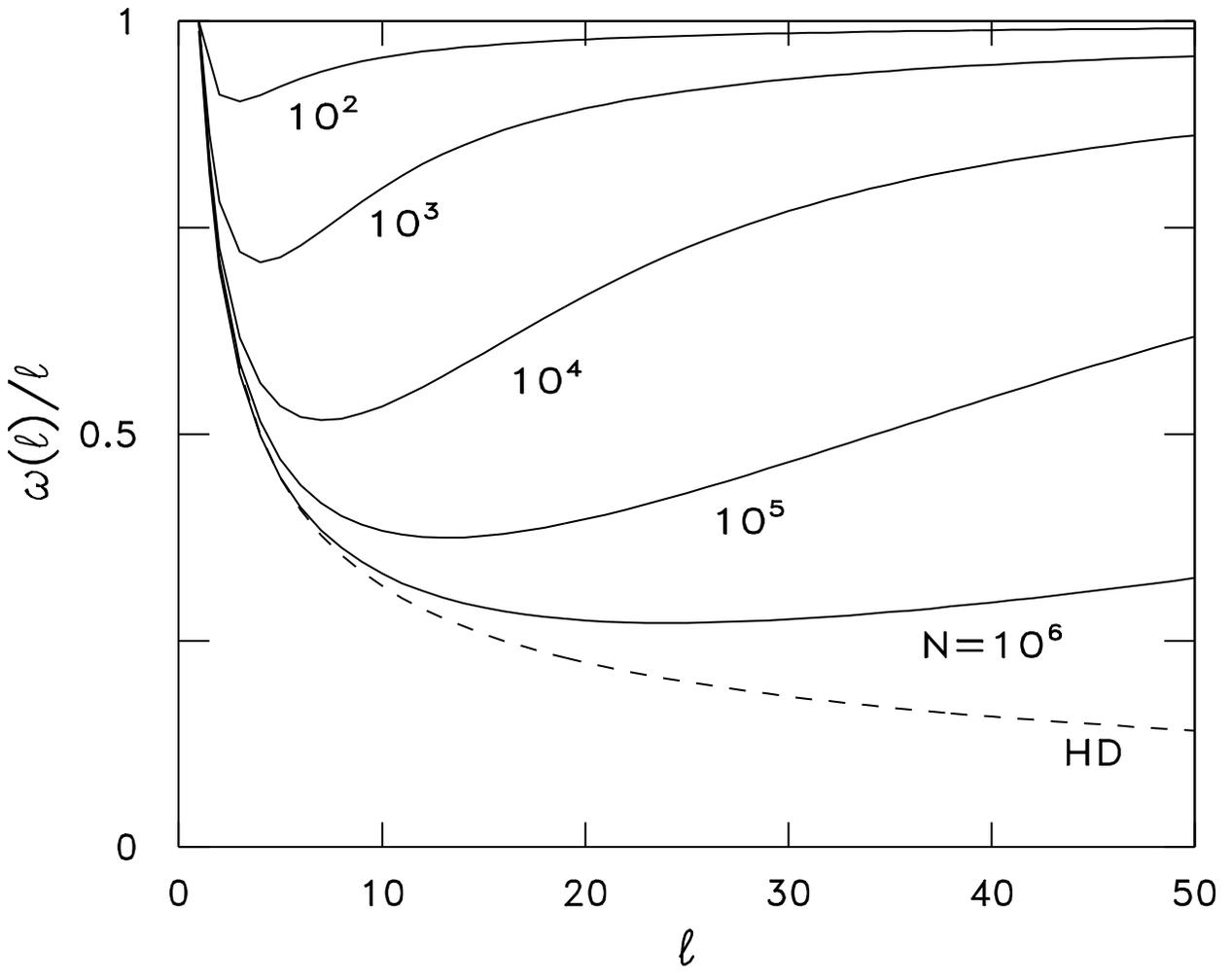
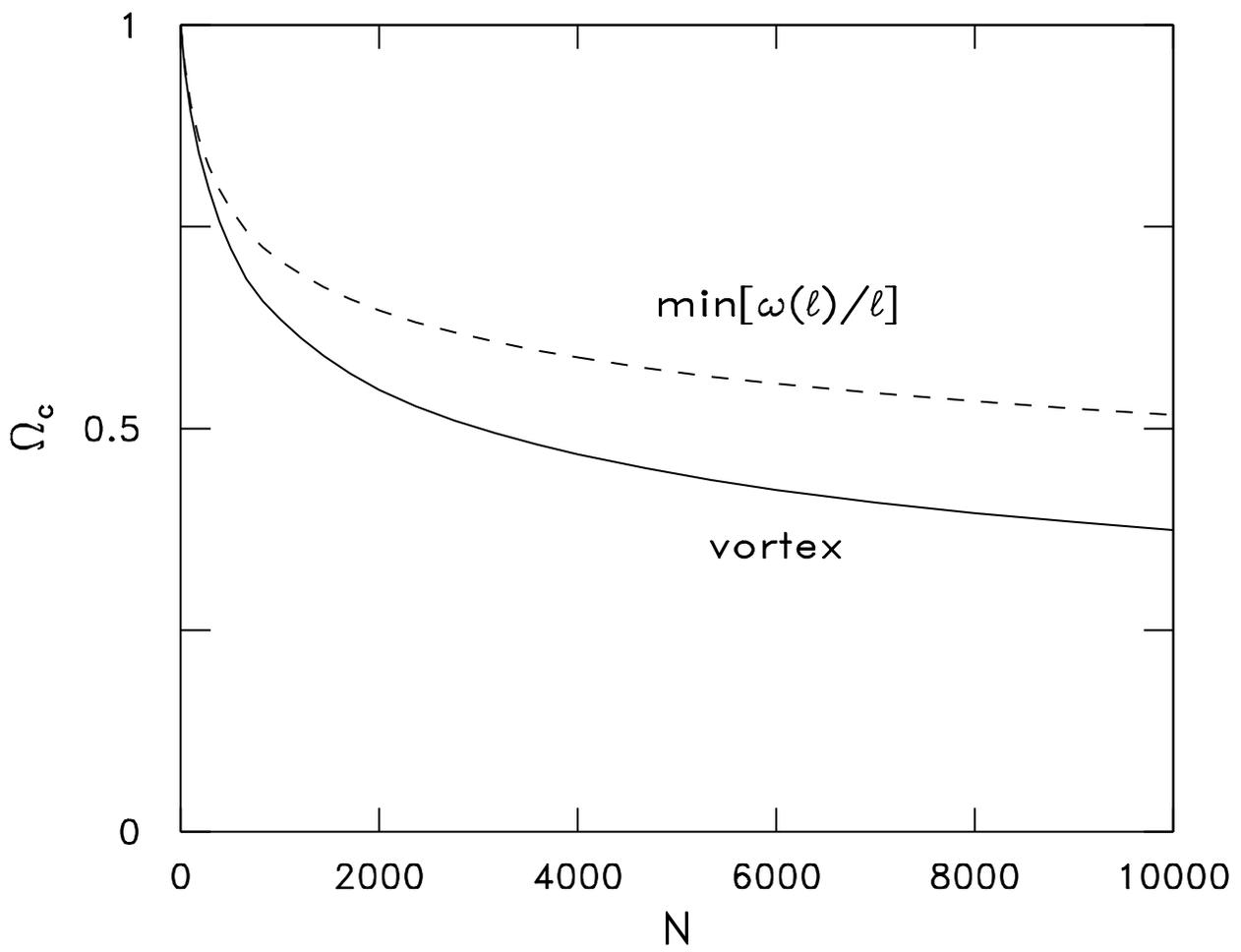
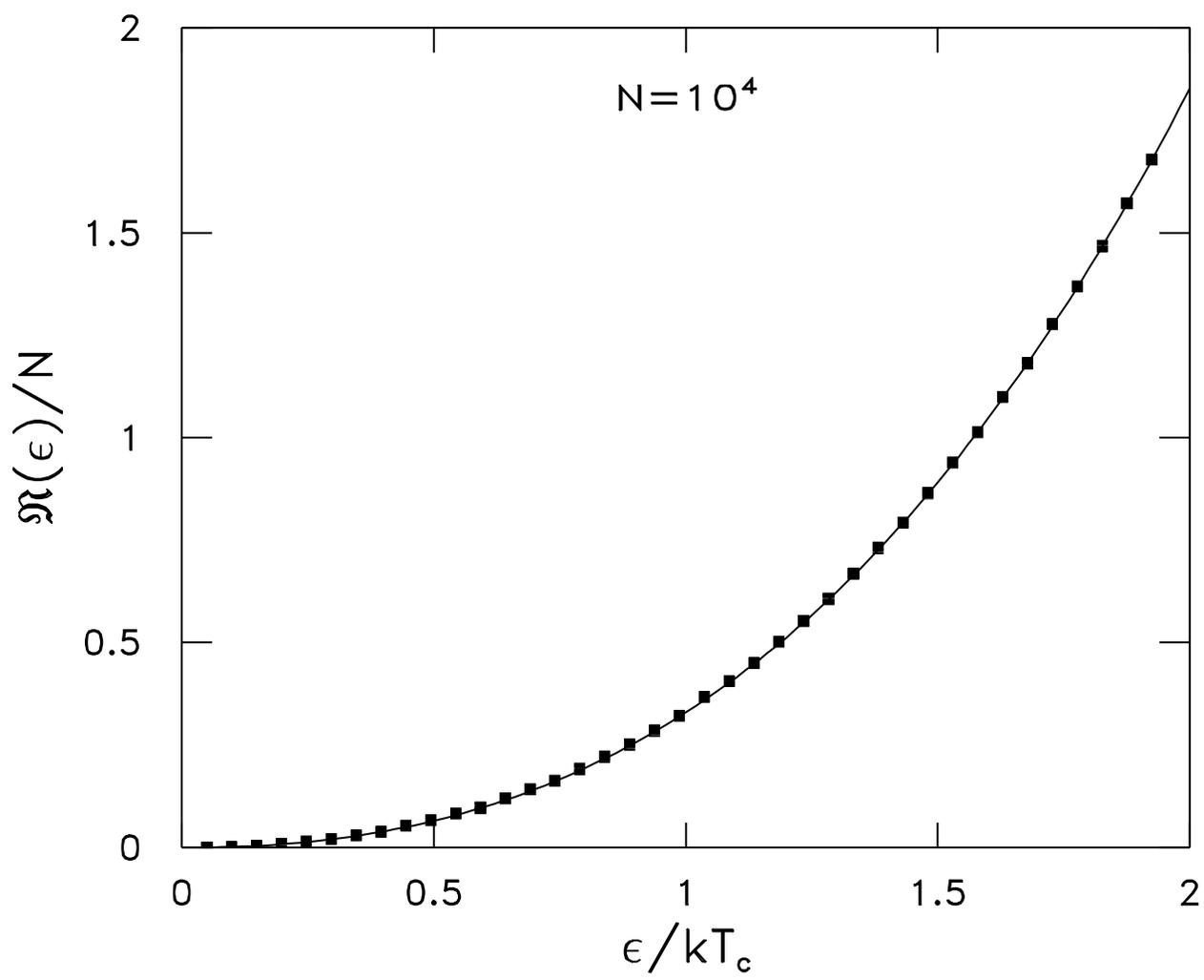
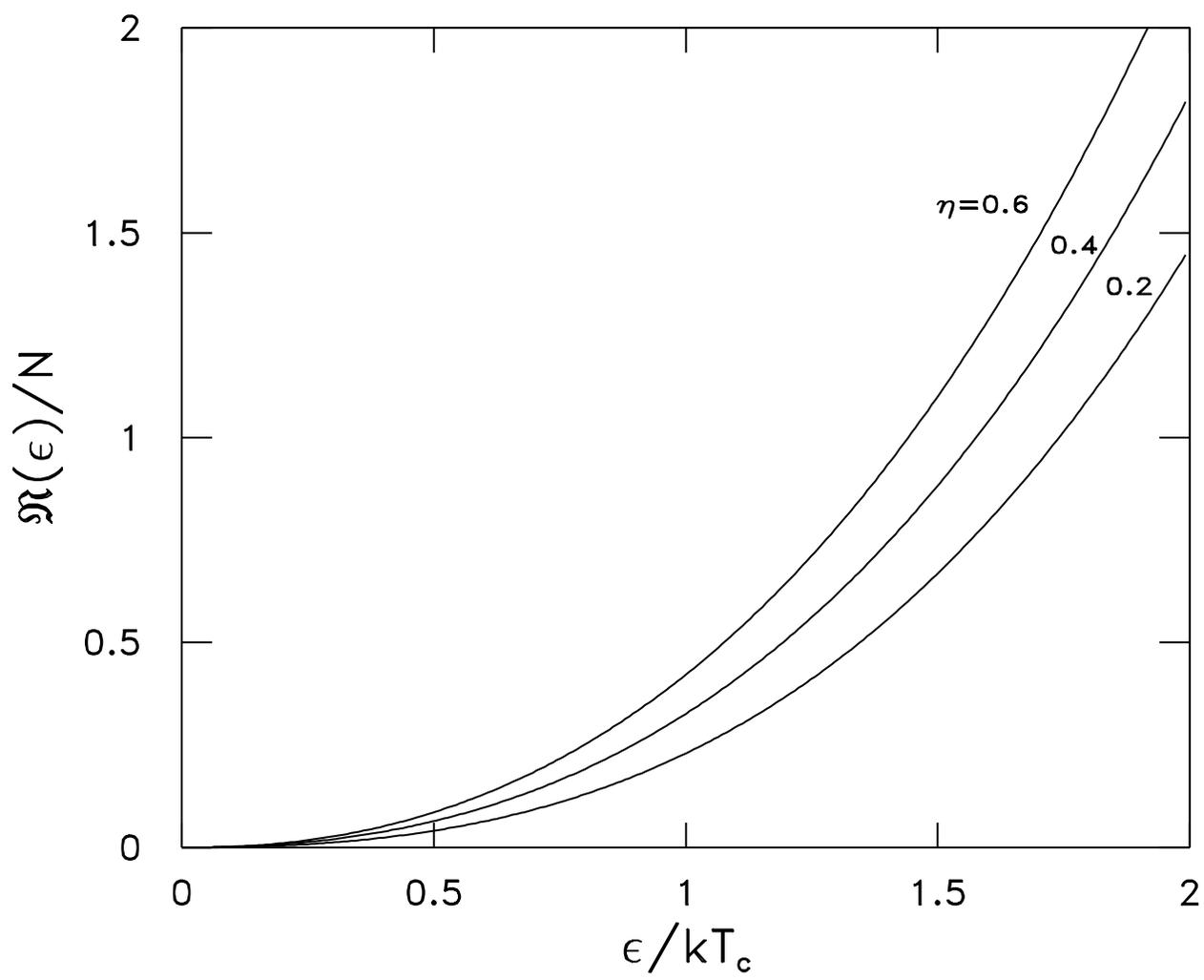
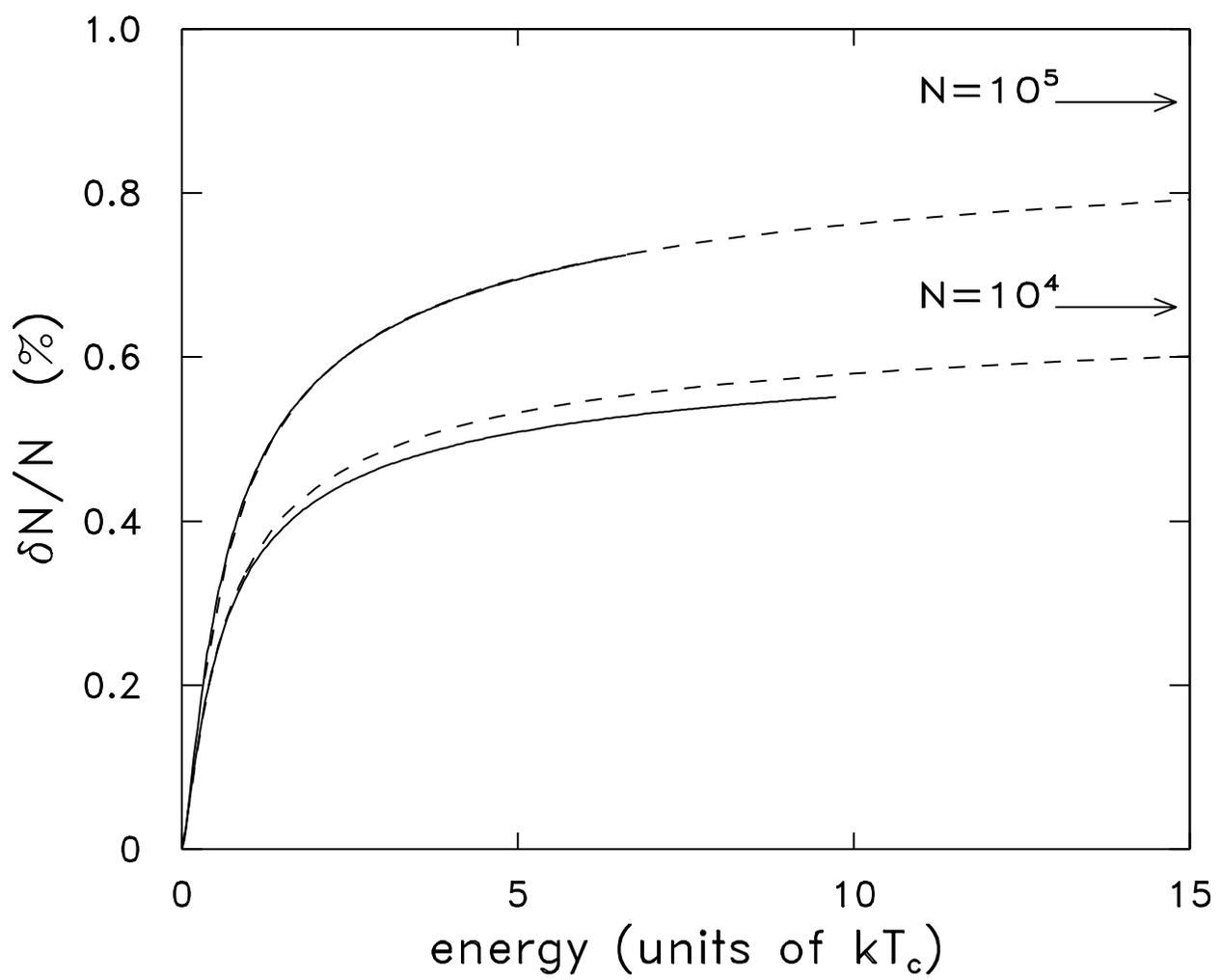
\end{document}